\documentclass[aps,superscriptaddress,11pt,showpacs,11p]{revtex4}
\usepackage{graphicx,epsf,amssymb,amsmath,latexsym}
\newcommand{\be}{\begin{eqnarray}}
\newcommand{\ee}{\end{eqnarray}}
\def\ben{\begin{equation}}
\def\een{\end{equation}}
\def\bena{\begin{eqnarray}}
\def\eena{\end{eqnarray}}
\def\non{\nonumber}

\begin{document}

\title{Asymptotically $AdS$ brane black holes}

\vspace{.3in}

\author{Christophe Galfard}
\email{C.Galfard@damtp.cam.ac.uk}
\affiliation{D.A.M.T.P., Centre for Mathematical Sciences,
University of Cambridge, Wilberforce road, Cambridge CB3
0WA, England}
\author{Cristiano Germani}
\email{C.Germani@damtp.cam.ac.uk}
\affiliation{D.A.M.T.P., Centre for Mathematical Sciences,
University of Cambridge, Wilberforce road, Cambridge CB3
0WA, England}
\affiliation{Yukawa Institute for Theoretical Physics,
Kyoto University\\ Oiwake-cho Kitashirakawa Sakyo, Kyoto
606-8502, Japan}
\author{Akihiro Ishibashi}
\email{akihiro@oscar.uchicago.edu}
\affiliation{Enrico Fermi Institute, University of Chicago \\
Chicago, IL 60637, USA}

\vskip.3in
\begin{abstract}
We study the possibility of having a static, asymptotically AdS black hole
localized on a braneworld with matter fields, within the framework of
the Randall and Sundrum scenario.
We attempt to look for such a brane black hole configuration by slicing
a given bulk spacetime and taking ${\Bbb Z}_2$ symmetry about
the slices. We find that such configurations are possible, and
as an explicit example, we provide a family of asymptotically AdS brane
black hole solutions for which both the bulk and brane metrics
are {\em regular} on and outside the black hole horizon
and brane matter fields are {\em realistic} in the sense
that the dominant energy condition is satisfied. We also find that
our braneworld models exhibit signature change inside the black hole
horizon.

\end{abstract}

\begin{flushright}
DAMTP-2005-117\\
YITP-05-072
\end{flushright}

\maketitle
%
%\newpage

\section{Introduction}

In \cite{RS}, using the ``squashing" effect a negative bulk cosmological
constant has on a four dimensional hypersurface, a brane, Randall and
Sundrum (RS) uncovered a  mechanism of dynamical localization of gravity.
All dimensions, including the one(s) we don't see, could now be infinitely
large and one would still recover, on the brane, the General Relativistic
and Newtonian limits of gravity at low energies \cite{GT,Shiromizu}.
It has also been shown that Standard Model matter fields can be constrained
on a brane \cite{gonzalo} and observations do not rule out braneworlds
as cosmological models (see \cite{maartcosmo} for a review). From
the astrophysical point of view, both numerical and analytical models of
stars have been found \cite{toby,maarger}.

\par
On the other hand, black holes are not well understood in the RS braneworld
scenario. The first attempt to find a static black hole solution on
the brane was developed in \cite{Chamblin} where the 4D Schwarzschild black
hole metric on the brane was embedded in a 5D bulk containing
an extended singularity: a black string.
Were our universe to be a brane in a higher
dimensional bulk, such a state of affair is not satisfactory: one might
indeed expect astrophysical black holes formed by collapsing matter
{\it on the brane} to be localized on (or at least very close to)
the brane
(see however \cite{Maa}.)
%%%
Study of a simple gravitational collapse model \cite{BGM}
on a RS braneworld indicates the difficulty of finding a static
vacuum black hole solution localized on a RS braneworld.
%%%
The difficulties in understanding black hole solutions arise
from the fact that in general, brane dynamics generate bulk Weyl
curvatures, which then backreact on the brane dynamics.
One is then left with the very hard task of solving equations
of motion for the coupled system of bulk and braneworld
with given suitable initial data.
Such a program has not been answered yet,
and even numerical approaches are still rather
approximative \cite{Tanaka1}.
To simplify the problem, the authors of \cite{Dadhich}
looked for analytic solutions to the projected Einstein equations
on the brane only and found an exact (Reissner-Nordstrom looking)
black hole solution. Other similar solutions were subsequently
found \cite{Casadio1}. These bulks' geometries are not known.

\par %%%
Under such circumstances, it is interesting to ask whether a brane on
which a 4D black hole is localized can be found
by looking for a slice that intersects a bulk black hole.
However, generalizing the work of \cite{Chamblin},
Kodama showed in \cite{Kodama} that brane solutions
with a black hole geometry
cannot be found as a slicing of a bulk with $G(D-2,k)$ symmetry,
if the brane is {\em vacuum} and {\em not} totally geodesic
\footnote{$D$ is the dimensionality of the bulk spacetime
and $k$ is the $D-2$ sectional curvature $k=\pm 1,\,0$;
a brane is called totally geodesic if its extrinsic
curvature is vanishing. }.
In other words if, for simplicity, one wants to keep studying slices
of bulks with $G(D-2,k)$ symmetry to find localized black holes in
the RS braneworld scenario---in which the brane is not totally
geodesic---one has to look for a {\it non-vacuum} brane.

\par %%%

Recently, an attempt to find a localized static but {\em non-vacuum}
brane black hole solution as a slice of a $G(D-2,k)$ bulk was made by
Seahra~\cite{Seahra}.
There, the bulk chosen was the Schwarzschild and Schwarzschild-AdS black
hole with spherical three-dimensional geometry, i.e., $G(3,k=1)$, and
branes were taken as a planar, asymptotically flat slice of these bulks.
Unfortunately these slicing turned out to produce naked singularities
with respect to the induced geometry, except when corresponding to
the equatorial plane of a bulk black hole, a special case of a totally
geodesic brane.

The aim of this paper is to find a regular RS braneworld on which
a static, spherically symmetric black hole surrounded by realistic matter
is localized, by slicing a fixed 5D black hole bulk spacetime.
The choice of slicing we will use is motivated by the AdS/CFT-inspired
``classical braneworld black hole" vs ``quantum black hole" duality
of \cite{Emparan} which states:
``The black hole solutions localized on the brane in the $AdS_{d+1}$
braneworld which are found by solving the classical bulk equations in
$AdS_{d+1}$ with the brane boundary conditions, correspond to
quantum-corrected black holes in $d$-dimensions, rather than classical ones"
(see also \cite{Tanaka2}). Since due to Hawking radiation, black holes in
asymptotically flat spacetimes are semi-classically unstable, such a duality
would explain the impossibility \cite{BGM} of finding a static exterior to
the Oppenheimer-Snyder collapse of a star in asymptotically flat RS
braneworlds (see also \cite{Casadio2}). However, asymptotically AdS
spacetimes allow (big enough) black holes to be in semi-classical
equilibrium \cite{HawkingPage} with their Hawking radiation. That is the
main motivation for turning our attention to the specific slices we study,
which are non-vacuum and asymptotically AdS (in a weak sense, see
\ref{Sect:IV}).
Encouraging results were already obtained in \cite{Emparan2}.
%\par %%%
%The main purpose of this paper is to
In this paper we will indeed show with explicit examples that it is
possible to construct a localized braneworld black hole surrounded by matter
that satisfies the dominant energy condition, when the braneworld is
asymptotically AdS (the case of an asymptotically AdS brane black hole
as a slice of a bulk black string was studied in \cite{char}).

\par %%%
The plan of the paper is as follows.
In the next section, we show that regular slices that
cross a bulk black hole horizon can be constructed
and we point out why the planar slices of \cite{Seahra} exhibit
a curvature singularity there.
In Sect.~\ref{Understanding:Slicing}, we fix our notations and define
a bulk slice which is an asymptotically AdS braneworld.
We then consider a simple one-parameter family of slices which correspond
to asymptotically vacuum and asymptotically AdS braneworlds
with black hole horizon, filled with matter satisfying the dominant
energy condition. Under our slicing ansatz, we find such braneworlds are
possible only for the bulk with three-dimensional spatial geometries
corresponding to $k=0$ and $k=-1$.
For the bulk spacetimes with spherical three-dimensional geometry ($k=1$),
our slicing define braneworld with matter violating the dominant energy
condition. We also find that some of our braneworlds exhibit
an intriguing property: signature change inside brane black holes.
In the conclusion, we summarize and discuss our results.

%--------------------------------------------------------

\section{Slices in $G(3,k)$ bulk}
\label{Sect:II}
\subsection{Bulk spacetime, slicing, and singularity condition}

Bearing the Schwarzschild-AdS bulk metric in mind, we consider
the following type of five-dimensional static metrics with $G(3,k)$
spatial symmetry
\be
\label{our metric}
{}^{(5)} ds^2=-A(r)dt^2+\frac{dr^2}{A(r)}+ \frac{r^2}{l^2} d\sigma_{(3)}^2 \,,
\ee
where $d\sigma_{(3)}^2 =\gamma_{ij}dx^i dx^j$ is the line element of a
3-dimensional space of constant curvature $k=0 , 1$ or $-1$,
\be
d\sigma_{(3)}^2 =
 dw^2+l^2\frac{\sin^2({\sqrt{k}w}/{l})}{\sqrt{k}^2}d\Omega^2_{(2)} \,,
\ee
with $d\Omega^2_{(2)}$ being the metric of unit two sphere.
For now, we do not impose the Einstein equations so that the function $A(r)$
is arbitrary.

In \cite{Kodama}, Kodama has shown that such bulk metrics cannot
embed a {\it vacuum}, non-totally geodesic brane which describes
a static black hole.
%%%
We would like, in the following, to relax the vacuum condition,
and see whether a regular black hole  can then be constructed on the brane.
As we progress, we will restrict the bulk metric (\ref{our metric}) to obtain
an asymptotically AdS braneworld with a horizon and positive energy
density matter as a slice of such a bulk. It is useful to introduce a new
radial coordinate
\be
\label{rho coordinate}
 \rho \equiv r \frac{\sin ({\sqrt{k}w}/{l})}{\sqrt{k}} \ .
\ee
Then,
\be
 \frac{dw}{l}=\frac{\rho}{r} \frac{d\log (\rho/r)
                   }{\sqrt{1-k \left({\rho}/{r}\right)^2 }} \,,
\label{dw}
\ee
and the five dimensional metric becomes
\be
\label{bulk}
{}^{(5)}ds^2 = -A(r)dt^2
+\left[\frac{1}{A(r)}+\frac{\rho^2/r^2}{1-k\frac{\rho^2}{r^2}}\right]dr^2
-2\frac{\rho /r}{1-k\frac{\rho^2}{r^2}}drd\rho
+\frac{d\rho^2}{1-k\frac{\rho^2}{r^2}}
+\rho^2d\Omega^2_{(2)} \,.
\ee
Looking for asymptotically AdS branes, we consider slices $r=r(\rho)$
of the bulk (\ref{bulk}).
The induced $4D$ metric on such a slice becomes
\be
\label{4}
{}^{(4)}ds^2 &=&
-A(\rho)dt^2+\left(\frac{{r'}^{2}}{A(\rho)}+B(\rho)\right)d\rho^2
+\rho^2d\Omega_{(2)}^2 \,,
\ee with
\be
\label{B}
B(\rho)&=&\frac{\left(1-\frac{\rho}{r} r'
\right)^2}{1-\frac{k\rho^2}{r^2}} \,,
\eena
where here and hereafter $r'\equiv dr/d\rho$.

We are interested in slices that cross a bulk event
horizon, so we assume the existence of at least a zero of the
function $A(r)$. Let $r_0$ be such that $A(r_0)=0$.
We will construct our brane so that there is a $\rho_0 \;(>0)$
which satisfies $r(\rho_0)=r_0$. For simplicity we restrict
our bulk (\ref{our metric}) to those for which $A(r)$ is at least $C^2$
except at $r=0$, and ${dA}/{dr}|_{r_0}\neq 0$ (simple zero at $r=r_0$,
that is, the horizon is non-degenerate).
Hence in particular, our bulk spacetimes can be singular only at $r=0$.
It is clear that $\rho=\rho_0$ is also a Killing horizon
with respect to the brane metric (\ref{4}).

The curvature scalars of the metric (\ref{4}) are all of the form:
\bena
\non R^\alpha{}_{\alpha} &=&
\frac{P_1}{2\rho^2 \left(1-k\frac{\rho^2}{r^2}\right)^2
           \Big[r'^2 + A(\rho) B(\rho) \Big]^2} \,, \\
\non R^{\alpha\beta}R_{\alpha\beta} &=&
\frac{P_2}{8\rho^4 \left(1-k\frac{\rho^2}{r^2}\right)^4
           \Big[ r'^2 + A(\rho)B(\rho) \Big]^4} \,, \\
     R^{\alpha\beta\gamma\delta}R_{\alpha\beta\gamma\delta} &=&
\frac{P_3}{8\rho^4 \left(1-k\frac{\rho^2}{r^2}\right)^4
           \Big[ r'^2 + A(\rho)B(\rho) \Big]^4} \,,
\label{denominator of curvature scalars}
\eena
where the numerators $P$s are polynomial of $A, r, \rho$ and their
first and second order derivatives.
Therefore, possible curvature singularities with respect to the brane metric
can happen whenever one of the $P$s blows up---which happens only when
$A$ diverges at $r=0$---or when the denominator of
(\ref{denominator of curvature scalars}) becomes zero,
that is for instance when
\bena
&\mbox{a/}& \ \ \left(1- k\frac{\rho^2}{r^2} \right)
                \Big[ r'^2 + A(\rho)B(\rho) \Big]^2=0 \,,
\label{curvature singularity conditions} \\
&\mbox{b/}& \ \  \rho=0 \,,  \\
&\mbox{c/}& \ \ r(\rho)=0 \,\,, \mbox{if}\
\lim_{r\rightarrow 0}|A(r)|=\infty \,.
\eena
In the slices we will consider below, all the curvature singularities
are of one of the above types (they are not mutually exclusive).
The case c/ corresponds to 5D bulk singularities.
The case b/ would occur even when our bulk spacetime is non-singular,
due to a non-trivial embedding. The case a/ also may occur as
an embedding singularity, but we need to consider this case with much
more care.
While singularities of types c/ and b/ are shown to be hidden inside
the horizon on the brane $\rho =\rho_0$, singularities of type a/
can occur on the brane's horizon $\rho = \rho_0$, hence can be
a naked singularity for braneworld observers.
In fact, from the induced metric (\ref{4}),
one can show that near the horizon ($\rho=\rho_0$)
the Ricci scalar on the brane is
\be
  R^{\alpha}{}_{\alpha}
\simeq \frac{\frac{dA}{dr}(r_0)B(\rho_0)}{2r'(\rho_0)^3} \,.
\ee
Therefore, in particular, for $B$ such that $B(\rho_0)\neq 0$,
there is an embedding curvature singularity where the brane crosses
the bulk event horizon if the slice is so that $r'(\rho_0)=0$.
We will see below that the naked singularities on the planar slices
in \cite{Seahra} are of this type.

\par %%%
It is intriguing to note that although the bulk metric~(\ref{our metric})
itself is everywhere Lorentzian, the induced metric (\ref{4}) can be
Euclidean inside the event horizon, $\rho < \rho_0$,
if $B(\rho)$ can take a sufficiently positive large value
so that the $(\rho,\rho)$ component of the brane metric, eq.~(\ref{4}),
can be positive, there.
If this is the case, such a brane displays a signature change
on a braneworld~\cite{Vera,AkiGary}.
It is interesting to notice that in Sect.~\ref{Sect:IV} we find such a
signature changing slicing under our requirements that branes
be asymptotically AdS, regular (at least on and outside the event horizon)
and that matter content satisfy the dominant energy condition.

%----------------------------------------------------------------------------------------------------------------------

\subsection{Comparison with earlier results from planar slicing}
\label{Comparison}

%----------------------------------------------------------------------------------------------------------------------

In \cite{Seahra}, Seahra has shown that planar slicing of 5D Schwarzschild
and Schwarzschild-AdS bulks are singular as soon as the slices cut the
bulk horizon ($r=r_0$). We will now prove that the planar slices
examined in \cite{Seahra} correspond to slices for which $r'=0$ on
the horizon, therefore corresponding to singularities of type $a/$.

\par
To find planar slices, introduce the following variable
\be
\label{Seahra}
R(r)= \exp \left( {\int^r_{r_1}} \frac{du}{\sqrt{u^2A(u)}} \right)\ ,
\ee
where $r_1$ is a fiducial initial distance, and define the cylindrical
coordinates as
\be
x=R\sin ({w}/{l}) \ ,\ y=R\cos ({w}/{l}) \ .
\label{coods:xy}
\ee
The bulk metric is then expressed as
\bena
 {}^{(5)}ds^2 = - A(r)dt^2 + \frac{r^2}{x^2+y^2}
          \left( dy^2 + dx^2 + x^2 d\Omega_{(2)}^2 \right)  \,,
\eena
where $r$ is now a function of $x^2+y^2$ through eqs.~(\ref{Seahra})
and (\ref{coods:xy}). We therefore find that planar slices
correspond to $y=const$, which we differentiate to obtain
an $r=r(w)$ expression for the slice.
(We put aside the special cases where $w$ is constant, which can always
be taken to correspond to $w/l =\pi/2$.)
Using eq.~(\ref{Seahra}), we have
\be
0 = \frac{d}{dw}\big[ R(r)\cos(w/l) \big]
 = R(r) \left[
               \frac{dr/dw}{\sqrt{r^2A(r)}}\cos(w/l) -\sin (w/l)
        \right] \,.
\ee
For non-degenerate horizons (i.e., $A$ has a single zero at $r=r_0$),
this implies
\be
\frac{dr}{dw}=r\sqrt{A(r)}\tan \left( {w}/{l}\right) \,,
\ee
everywhere. In particular, ${dr}/{dw}\Big|_{r_0}=0$.
On the other hand, using the coordinate $\rho$ introduced in
eq. (\ref{rho coordinate}), we have
\be
\frac{dr}{dw} =
 \frac{\sqrt{1 - k({\rho}/{r})^2}}{1- ({\rho}/{r})(dr/d\rho)}
 \frac{r}{l} \frac{dr}{d\rho} \,.
\ee
If for planar slices, $r \neq \rho$ at the horizon,
then $dr/dw\Big|_{r_0}=0$ implies $dr/d\rho \Big|_{r_0}=0$.
We showed above that such slices possesses an embedding singularity
at the horizon, singularity which was analyzed in \cite{Seahra}.
If $\rho=r$, $w/l=\pi/2$ ($k=+1$) (see eq. (\ref{rho coordinate}))
so that $dr/d\rho = 1 \neq 0$, this corresponds to a totally geodesic
brane, equatorial slicing of a Schwarzschild or Schwarzschild-AdS
bulk black hole.

%---------------------------------------------------------------------------

\section{Understanding the slice $r(\rho)$ as a brane world}
\label{Understanding:Slicing}

%---------------------------------------------------------------------------

In order to establish our notations, we here recall some
basic equations used in the RS braneworld scenario, in particular
the relation between the extrinsic curvature of a slicing $\Sigma$
and energy-momentum tensor for matter fields confined on $\Sigma$.
%%%
For simplicity, we assume that our 5D bulk metric $g_{\mu \nu}$ obeys
the vacuum Einstein equations with a negative cosmological constant
$\Lambda_5$.
Assigning surface intrinsic energy momentum tensor $T_{\mu \nu}$
and a surface tension to $\Sigma$, we now think of $\Sigma$ as
a gravitating brane, hence our basic equation is
\be\label{5d}
{}^{(5)}G_{\mu \nu}=-\Lambda_5 g_{\mu \nu}+\kappa_5^2
  \left(
        T_{\mu \nu} - \frac{6}{\kappa_{5}^2} \sigma q_{\mu \nu}
  \right) \delta(\chi) \,,
\ee
where ${\kappa_5^2}$ denotes 5D gravitational constant,
and the constant $\sigma$ is in proportion to the surface tension.
Here $q_{\mu \nu}$ is the projection tensor (also referred to
interchangeably as the induced/brane metric) for $\Sigma$ defined in terms of
the unit normal vector $n^\mu$ by $q_{\mu \nu} = g_{\mu \nu} -n_\mu n_\nu$,
and the function $\chi$ is introduced to specify the location of $\Sigma$
as $\chi=0$ and $n^\mu \partial_\mu \chi = const$ on $\Sigma$.

We define the extrinsic curvature of $\Sigma$ by
\ben
\label{K}
 K_{\mu \nu} = s \frac{1}{2}{\cal L}_n q_{\mu \nu} \,,
\een
where $s = \pm 1$ is introduced for later convenience.
%%%
Since $\Sigma$ is a co-dimension one surface, it divides our bulk
spacetime into two regions. The `orientation' $s$ decides
which side of the bulk, we will consider as the ambient spacetime of
$\Sigma$.
%%%
We further assume ${\Bbb Z}_2$ symmetry with respect to $\Sigma$ and
take the normal $n^\mu$ so that when $s=+1$, $n^\mu$ is
directed toward inside the ${\Bbb Z}_2$ symmetric bulk.
Then, integration of eq.~(\ref{5d}) along $\chi$ or the junction
condition yields
\be
\label{Tsigma}
T^\mu{}_\nu = \frac{2}{\kappa_5^2}
              \Big(
                    - K^\mu{}_\nu + (K+3\sigma)\delta^\mu{}_\nu
              \Big) \,.
\ee

In order to understand the effective gravitational theory induced
on the brane it is useful to consider 4D projected components
of eq.~(\ref{5d})~\cite{Shiromizu,maartcosmo},
\be
\label{G}
{}^{(4)}G_{\mu\nu}=-\Lambda_4 q_{\mu\nu}+{\cal T}_{\mu\nu} \,,
\ee
where ${}^{(4)}G_{\mu\nu}$ is the four dimensional Einstein tensor
on the brane and ${\cal T}_{\mu\nu}$ is the effective energy momentum
tensor as seen by a brane observer,
\be
\label{Teff}
{\cal T}_{\mu\nu}= \sigma \kappa_5^2T_{\mu\nu}
                 + {\kappa_5^4}{\cal S}_{\mu\nu}
                 - {\cal E}_{\mu\nu} \,,
\ee
where ${\cal E}_{\mu \nu}$ is the projected 5D Weyl tensor, and
\bena
{\cal S}_{\mu\nu} &=&
    \frac{1}{12}TT_{\mu\nu}
    -\frac{1}{4}T_{\mu\gamma}T^{\gamma}{}_{\nu}
    +\frac{1}{24} q_{\mu\nu}
                 \left(3T_{\gamma\delta}T^{\gamma\delta}- T^2\right) \ , \\
\Lambda_4&=&\frac{1}{2}\left(\Lambda_5+6\sigma^2\right)\,.
\label{L4}
\ee
We will denote {\it real} energy-density and pressures coming from
$T^\mu{}_\nu$ in eq.~(\ref{Tsigma}) by
\bena
  T^t{}_t = - \epsilon \,, \quad
  T^i{}_i = p_i \,,
\label{real}
\eena
and the {\it effective} energy density and pressures coming from
${\cal T}^\mu{}_\nu$ in eq.~(\ref{Teff}) by
\bena
{\cal T}^t{}_t = - \tilde\epsilon \,, \quad
{\cal T}^i{}_i =   \tilde p_i \,.
\label {effective}
\eena

%---------------------------------------------------------------------------
\section {Localized brane black holes in a Schwarzschild-AdS bulk}
\label{Sect:IV}
%---------------------------------------------------------------------------

In this section, as our $G(3,k)$ symmetric bulk spacetime
we shall consider 5D Schwarzschild-AdS type metrics
(\ref{our metric}) with
\be
A(r)=k-\frac{2Ml}{r^2}+\frac{r^2}{l^2}\,,
\ee
which are solutions to 5D vacuum Einstein equations with
a negative constant $\Lambda_5$ with $l^2=6/|\Lambda_5|$
and $M$ is related to the mass of the bulk black hole.
Note also that for geometries $k=0$ and $k=-1$, the spatial section of
bulk Killing horizons are not compact unless a suitable identification
of points along space-like directions is made.

We are looking for a regular asymptotically AdS brane which possesses
an event horizon surrounded by matter satisfying some desirable energy
conditions (the dominant one).
In this paper, we call our brane asymptotically AdS
if the asymptotic expansion in $\rho$ of the first and second fundamental
forms of the brane match, to leading order, the asymptotic expansion of
a pure AdS brane.
In terms of the coordinate system we are employing,
our slices are asymptotically AdS of the above sense as soon as
\bena
 q_{tt} \sim \frac{\rho^2}{L^2} + O(1) \,, \quad
 q_{\rho \rho} \sim
 \frac{L^2}{\rho^2} + O\left(\frac{L^4}{\rho^4}\right) \,, \quad
\rho \rightarrow \infty \,,
\eena
and $-3/L^2=\Lambda_4$ where $\Lambda_4$ is defined in eq.~(\ref{L4}).
The condition on the second fundamental form is then immediately
satisfied. This asymptotic AdS condition is slightly different
and weaker than the asymptotic AdS conditions of \cite{Henneaux},
where the metric is required to have asymptotic symmetry
group $SO(2,3)$ in 4 dimensions.

\par %%%
A particularly interesting class of slicing that can be
asymptotically AdS in the above sense is given by
\be
 r=\rho+ \gamma \, ,
\label{condi:slice}
\ee
where $\gamma$ is a constant parameter.
We assume that $\gamma > - \rho_0$.
In the rest of the paper we focus on braneworlds given
by the above slicing condition (\ref{condi:slice}).
The effective cosmological constant on such branes
will be related to this parameter.

%------------------------------------------------------------------------------

\subsection{Asymptotic energy conditions and singularities}

Now that we know what the condition for our slices not to be singular
where it crosses the bulk horizon is $r'(\rho_0)\neq0$, we would like to
see what are the asymptotic conditions for the slice in order for its matter
content not to violate the dominant energy condition near spatial infinity.
The dominant energy condition requires energy density to be positive
and the absolute value of the radial and angular pressures to be smaller
than the energy density.

%%%
We first derive an expression for the asymptotic form of
the {\it real} energy density $\epsilon$ and pressures $p_i$
for a braneworld defined by (\ref{condi:slice}).
%%%
Defining the unit normal vector to our slicing $r=\rho + \gamma$ as
\be
n^\mu = g^{\mu \nu}N \left(\partial_\nu r - r' \partial_\nu \rho \right)
      \,, \,\, \mbox{where}\,\,
N= \frac{1}{\sqrt{A(1-r'\rho/r)^2 + r'^2(1-k\rho^2/r^2)}} \,,
% \frac{1}{\sqrt{g^{rr}-2g^{r\rho}r'+g^{\rho\rho}r'{}^2}} \,,
\label{normal}
\ee
%%%
and using the formulae~(\ref{4}), (\ref{K}), (\ref{Tsigma}) and (\ref{real})
given in the previous sections, we can calculate $\epsilon$ and $p_i$.
Since the explicit expressions of the extrinsic curvature $K^\mu{}_\nu$
and the stress energy tensor $T^\mu{}_\nu$ themselves are rather complicated
and not so illuminating, we do not present them here.
The results give, asymptotically,
\bena
\label{real T with sigma}
\epsilon &=&
-\frac{3(s\gamma+\sigma l \sqrt{\gamma^2+l^2(1-k)})}{l\sqrt{\gamma^2+l^2(1-k)}}
+\frac{2s(\gamma^2+l^2(1-k)^2)}{\left(\gamma^2+l^2(1-k) \right)^{3/2}}
      \frac{l}{\rho}+O\left(\frac{l^2}{\rho^2}\right)\,, \\
p_\rho &=&
\frac{3(s\gamma+\sigma l \sqrt{\gamma^2+l^2(1-k)})}{l\sqrt{\gamma^2+l^2(1-k)}}
-\frac{s(\gamma^2(2+k)+2l^2(1-k)^2)}{\left(\gamma^2+l^2(1-k) \right)^{3/2}}
 \frac{l}{\rho}+O\left(\frac{l^2}{\rho^2}\right) \,, \\
p_\theta &=&
\frac{3(s\gamma+\sigma l \sqrt{\gamma^2+l^2(1-k)})}{l\sqrt{\gamma^2+l^2(1-k)}}
-\frac{s(\gamma^2(1+k)+l^2(1-k))}{\left(\gamma^2+l^2(1-k) \right)^{3/2}}
 \frac{l}{\rho} + O\left(\frac{l^2}{\rho^2}\right) \,.
\eena
%%%
It is to be noted that the parameter $M$, which is proportional to the mass
of bulk black hole, plays no role until the next order (see for instance
eq.~(\ref{next order}) below).

\par %%%
In order to have an asymptotically empty brane, we fine-tune,
\`a la Randall-Sundrum, the vacuum energy related to $\sigma$
\be
\label{sigma fine tuning}
\sigma = \sigma_0 \equiv -\frac{s\gamma}{l\sqrt{\gamma^2+l^2(1-k)}} .
\ee
The constant term in the above expressions then vanishes.
We will keep such a fine-tuning throughout.
It is straightforward to see that for the choice $s=+1$, the energy density
is asymptotically positive for $k=\pm1$ and $k=0$.
For the energy density to pressures ratios, we asymptotically find:
\bena
\left|\frac{\epsilon}{p_\rho}\right|
 &=&
\left|\frac{2(\gamma^2+l^2(1-k)^2)}{\gamma^2(2+k)+2l^2(1-k)^2}\right|
  + O\left( \frac{l}{\rho} \right) \ , \\
\left|\frac{\epsilon}{p_\theta}\right|
 &=& \left|\frac{2(\gamma^2+l^2(1-k)^2)}{\gamma^2(1+k)+l^2(1-k)^2}\right|
  + O\left( \frac{l}{\rho} \right) \ .
\ee
So,
\bena
\left|\frac{\epsilon}{p_\rho}\right| \ \rightarrow \
\begin{array}{ccc}
& >1 & \ \mbox{for} \ k=-1\ \mbox{and} \ \gamma \neq 0 \,, \\
& =1 & \ \mbox{for} \ (k=0)\  \mbox{or}
                         \ (\gamma = 0 \ \mbox{and} \ k\neq 0) \,, \\
& = 2/3<1 & \ \mbox{for} \ k = 1 \,.
\end{array}
\eena
For the dominant energy condition to be satisfied asymptotically,
we thus have to rule out the $k=1$ case. It is easy to see that
$|\epsilon|/|p_\theta| >1$, %%% \footnote{
except when $\gamma=0$ and $k=1$,
which is a totally geodesic brane ($K_{\alpha\beta} = 0$).
We will not consider that case any further.
The $ (k=0)\  $ case requires a next order analysis, and we find
\be
\label{next order}
\left|\frac{\epsilon}{p_\rho}\right|
 = 1 - \frac{2\gamma^3 M}{l^2(\gamma^2+l^2)}\frac{l^3}{\rho^3}
 + O \left(\frac{l^4}{\rho^4} \right) \,,
\ee
so that in the $k=0$ case, the dominant energy condition (DEC) is
only satisfied asymptotically if $\gamma <0$
(corresponding to a brane with positive vacuum energy).

Recapitulating (and completing) the asymptotic behavior of
the real matter, we find (``$Yes$" means DEC satisfied asymptotically and
``$No$" means DEC violated)
\be
 \begin{array}{ccc}
 & &  \\
& \gamma >0 &  \\
& \gamma=0 & \\
& \gamma<0 & \\
  \end{array}
  \begin{array}{ccc}
 & k=-1&  \\
& Yes &  \\
& Yes & \\
& Yes & \\
  \end{array}
   \begin{array}{ccc}
 & k=0&  \\
&  No &  \\
&  Yes & \\
&  Yes & \\
  \end{array}
  \begin{array}{ccc}
& k=1&  \\
& No &  \\
&  - & \\
& No & \,. \\
  \end{array}
\ee
Note that the $k=1$, $\gamma=0$ case %%% cannot be obtained by going from
%%% eq.(\ref{rho coordinate}) to eq.(\ref{dw}) since it
corresponds to a $w/l = \pi/2$, $r=\rho$ slice.
For such a slice, it is straightforward to obtain the induced metric
from (\ref{our metric}), and it corresponds to a totally geodesic brane.

\par
\begin{figure}[t]
\centering\includegraphics[angle=0,width=4.5in]{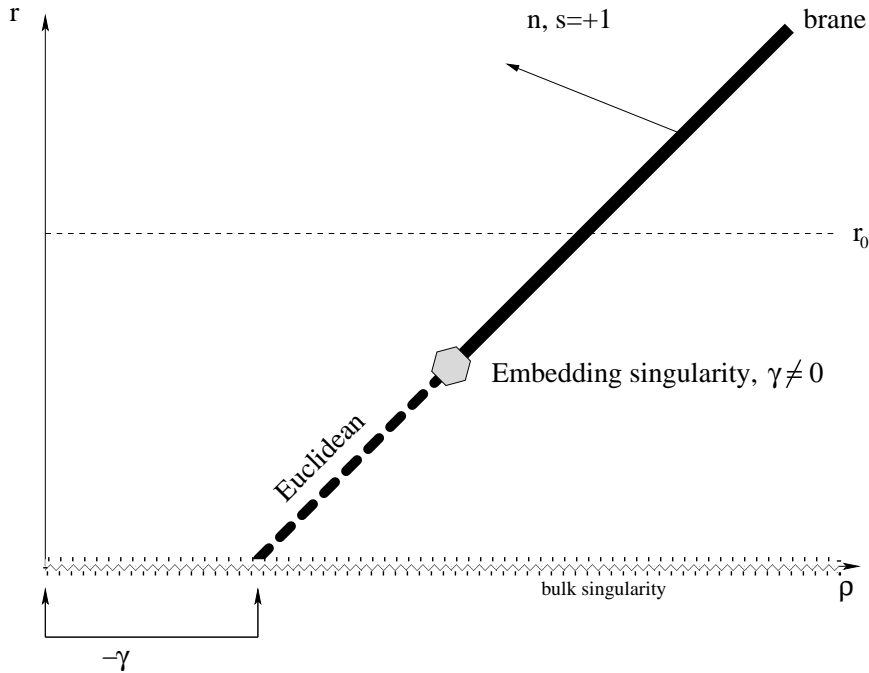}
\caption{
         Schematic representation of the slice $r=\rho+\gamma$
         in the $k=0,-1$ bulk. For $s=+1$, the upper side of
         the thick line (brane) corresponds to our bulk under
         ${\Bbb Z}_2$ symmetry. Our brane cut off a compact portion
         from the bulk event horizon (dashed horizontal line).
        Note that the 2-dimensional subspace spanned by $(r,\rho)$ is Riemannian above the line $r=r_0$ but is Lorentzian below the line. \label{brana}}
\end{figure}

 %%%
Thus, from the above observations of braneworld stress-energy,
physically interesting cases are those in which the bulk has a spatial
geometry of either $k=0$ or $k=-1$, and a slicing $\Sigma$ is
given by eq.~(\ref{condi:slice}) with $\gamma \leq 0$, and the bulk
orientation is $s=+1$.
%%%
We note that for $k=0$ and $k=-1$, the bulk event horizon is no longer
spatially compact. On the other hand, our spherically symmetric slicing
$\Sigma$ intersects the bulk event horizon at $\rho = \rho_0$,
and the region $\rho > \rho_0$ of $\Sigma$ always is outside the bulk
event horizon.
Therefore $\Sigma$ divides the bulk black hole horizon into two parts:
(i) a spatially compact portion,
$\{ r=r_0 \,,\,\, 0\leq \rho \leq \rho_0 \}$,
which is spherically symmetric, and (ii) its complement,
$\{r=r_0\,,\,\,\rho_0 <\rho < \infty \}$, which is infinitely extended
in the bulk. Then, depending upon the choice of $s=\pm 1$
(i.e., which side of the bulk we discard before taking ${\Bbb Z}_2$
symmetry), the resultant ${\Bbb Z}_2$ symmetric bulk will contain
either a spatially compact portion of the bulk event horizon,
or an infinitely extended portion of the bulk horizon.
Our choice $s=+1$ is indeed the former case.
In fact, noting that the coordinate components of the normal vector,
eq.~(\ref{normal}), are
\ben
 n^r = -NA\frac{-\gamma}{r} \,, \quad
 n^\rho = - NA\left\{
                     \frac{-\gamma \rho}{r^2}
                    +\frac{1}{A}\left(1-k \frac{\rho^2}{r^2}\right)
            \right\} \,,
\een
one can easily see that $n^\rho < n^r$ outside the event horizon
and hence that $n^\mu$ is directed toward the region
$\{r > \rho + \gamma \}$, as depicted in fig.~(\ref{brana}).
With the choice $s=+1$, this region corresponds to our ${\Bbb Z}_2$
symmetric bulk spacetime.
%%%
% eq.~(\ref{normal}), and $sn^\rho<0$ outside the horizon,
% as in our case $r'>0$ and $g^{r\rho}=-\rho/(rg^{tt})$ $sn^\rho<0$
% for $s=+1$.
%%%
We would like to emphasize that this choice $s=+1$ is the case
we consider in order to satisfy the energy condition for real matter.
By this construction, the resultant ${\Bbb Z}_2$ symmetric 5D spacetime
contains a RS brane as a boundary and a spatially compact portion
of the 5D event horizon attached to the brane.
This black hole looks like a compact object from both the bulk metric
and the brane's intrinsic metric view points, hence one can interpret
this geometry as a black hole localized on the brane.
%We note that when $\gamma\neq 0$ the singularity inside
%this black hole appears to occur at finite radius $\rho = -\gamma > 0$
%from the braneworld view point. This singularity shrinks to a
%point whenever $\gamma=0$.
%For $\gamma>0$ the brane can avoid the bulk singularity.

\par %%%

Our braneworld may change the signature of its induced metric,
despite the fact that the bulk spacetime is everywhere Lorentzian.
In fact, by inspecting $B(\rho)$ of eq.~(\ref{B}), we find that
for $k=-1,\,0$ whenever $\gamma \neq 0$, our braneworlds display
such a signature change inside the event horizon.
When $\gamma =0$, no signature change happens for any $k$.
(Note that for $k=+1$, the signature change can occur inside
the horizon if $\gamma >0$, but if $\gamma <0$, it can occur
even outside the horizon, at large $\rho $.)
%%%
Furthermore, we can see that the brane's intrinsic geometry
becomes singular at signature changing points. Such signature changes
have been studied e.g. in \cite{AkiGary}. Indeed, we find
from eq.~(\ref{4}) that in general the braneworld signature change
can occur when
$q_{tt} q_{\rho \rho} = - \left\{r'^2 + A(\rho)B(\rho) \right\}$
changes its sign, where $q_{\mu \nu}$ is the induced metric, eq.~(\ref{4}).
It then immediately follows from (\ref{curvature singularity conditions})
that the associated singularities are of type a/.
% This is all in agreement with the general analysis made
% at the beginning of the paper.

%-----------------------------------------------------------------------

\subsection{Black hole on the brane}
\label{subsect:singnature change}

%-----------------------------------------------------------------------

Here we study the character of the stress energy on our braneworlds,
$\epsilon$ and $p_i$, in more detail.
As a typical case we focus on
the $\gamma<0$ case. The analysis of the dominant energy condition above
restricts us to the $k=-1$ case. The induced 4D metric is
\bena
\label{induced gamma < 0 case}
{}^{(4)}ds^2 &=&
- \left(-1-\frac{2Ml}{(\rho+\gamma)^2}
+ \frac{(\rho+\gamma)^2}{l^2}\right)dt^2
+ \left[\frac{1}{-1-\frac{2Ml}{(\rho+\gamma)^2}
       +\frac{(\rho+\gamma)^2}{l^2}}
       +\frac{\gamma^2}{(\rho+\gamma)^2+\rho^2}
  \right] d\rho^2
 + \rho^2 d \Omega_{(2)}^2 \,.
\eena
One can immediately see that there is an event horizon on the brane
where the brane intersects the bulk event horizon. We now have a look
at what this horizon hides. The general analysis in Sect.~\ref{Sect:II}
shows that in the $k=-1$ case we are considering here,
since $r'=1$ and $A(\rho)$ has no zero outside the horizon, there can
be no curvature singularity outside the horizon. Indeed, the Ricci,
Ricci square and Riemann square curvature scalars for the induced
metric (\ref{induced gamma < 0 case}) can be written as before
eq.~(\ref{denominator of curvature scalars}).
%For the Ricci scalar, its denominator is
%\be
%Q\propto
%  \rho^2 (\rho+\gamma)^2
%   \left[ 1 + \frac{\rho^2}{(\rho+\gamma)^2}
%         +\left(-1-\frac{2Ml}{(\rho+\gamma)^2}
%         +\frac{(\rho+\gamma)^2}{l^2}\right)
%          \left(1-\frac{\rho}{\rho+\gamma} \right)^2
%   \right]^2 \,,
%\ee
%and similarly for the other two
%(see eq.~(\ref{denominator of curvature scalars})).
Explicitly, curvature
singularities of the 4D metric (\ref{induced gamma < 0 case}) occur when
\be
&&
\mbox{a/} \ \  -1-\frac{2Ml}{(\rho+\gamma)^2}
 + \frac{(\rho+\gamma)^2}{l^2} = -\frac{(\rho+\gamma)^2+\rho^2}{\gamma^2} \ ;
\\
%%% which may only happen inside the horizon;
&&
\mbox{b/} \ \  \rho=0 \ ; \quad \mbox{or} \,, \\
&&
\mbox{c/} \ \  \rho=-\gamma \ .
\ee
However, since the bulk black hole horizon, from the brane point of view
is located at $\rho=\rho_0$, where
\be
\rho_0=\sqrt{l^2/2+\sqrt{l^4+8Ml^3}/2}-\gamma \,,
\ee
the curvature singularities on the brane are hidden by the horizon
for all the cases a/, b/, c/ .

At infinity, the induced metric can be expanded in $\rho/l$, giving
\be
\label{asympt AdS brane}
 q_{tt}&\sim& -\frac{\rho^2}{l^2}\,, \quad
 q_{\rho\rho} \sim \frac{l^2+\gamma^2/2}{\rho^2} \,.
\ee
Re-scaling the time coordinate %%% $t \rightarrow \frac{L^2}{l^2}t$,
$t \rightarrow \frac{l}{L}t$ with $L\equiv \sqrt{l^2 + \gamma^2/2}$,
one can check that (\ref{asympt AdS brane}) has the same asymptotic
behavior at the brane's infinity $\rho \rightarrow \infty$
as the $AdS_4$ metric with cosmological constant
$\Lambda_4\equiv-3/L^2$ corresponding to eq.~(\ref{L4}), that is
\be
\Lambda_4=-\frac{6}{2l^2+\gamma^2} \,.
\ee
One can also see the above result from the expansion of the Ricci scalar
\be
 R^\alpha{}_\alpha
 = -\frac{3}{L^2}-\frac{6\gamma(8l^2+\gamma^2)}{L(2l^2+\gamma^2)^2}
    \frac{L}{\rho}
   + O \left(\frac{L^2}{\rho^2} \right)\,.
\ee

Although in principle we have all the ingredients to obtain analytically
all the characteristics of these black hole solutions on the brane,
in the case of asymptotic non-totally geodesic brane $\sigma\neq 0$,
the expression for the stress tensor and the curvatures are quite involved.
We therefore verify numerically that many of our slices are explicit examples
of a regular brane localized
black hole with surrounding matter that fulfill the dominant energy
conditions {\it everywhere} outside the brane's Killing horizon.
We here set $l=1$ and consider dimensionless variables.

\begin{figure}[t]
\centering\includegraphics[angle=0,width=4in,height=4in]{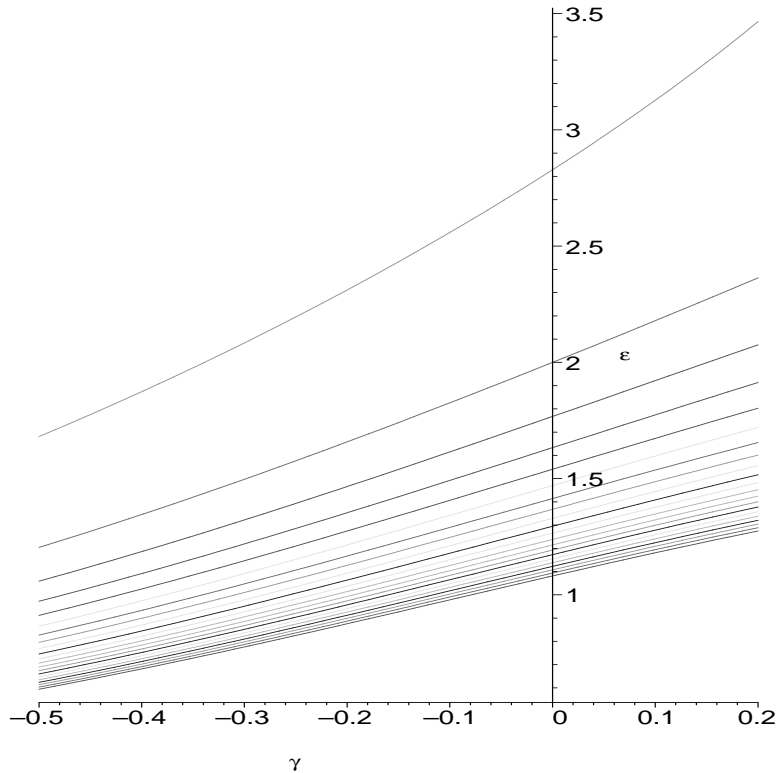}
\caption{
         Values of the real energy density at the horizon for
         $\gamma \in [-0.5,0.2]$ for $M=0...20$.
         The upmost plot corresponds to $M=0$, the lowest one to $M=20$.
         \label{WECh1}}
\end{figure}

\begin{figure}[t]\centering\includegraphics[angle=0,width=4in,height=4in]{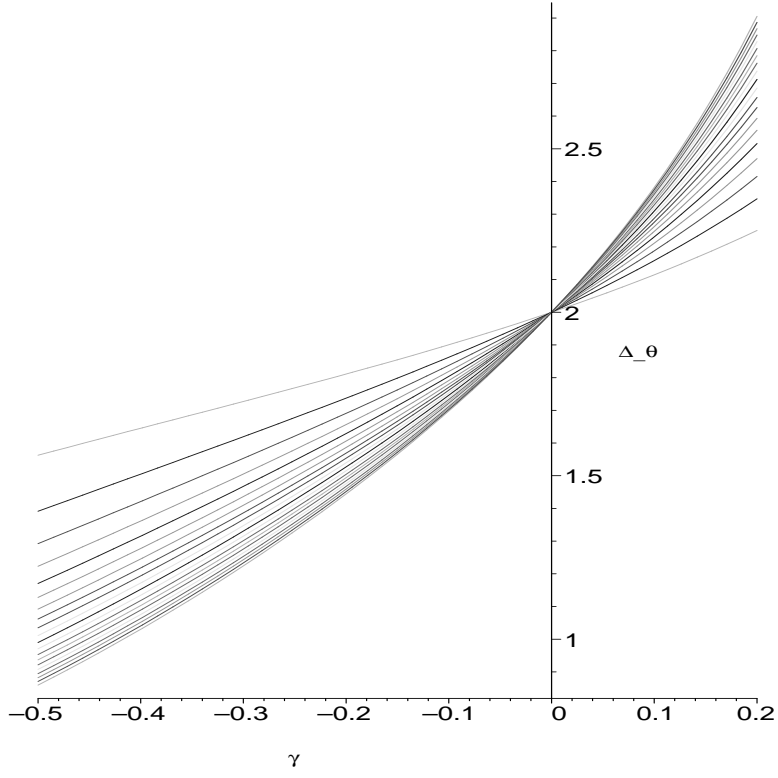}
    \caption{Values of $\Delta_\theta=|\epsilon/p_\theta|$  at the horizon for
             $\gamma \in [-0.5,\;0]$ for $M=1...20$.
 \label{DECh3}}
\end{figure}

In fig.~(\ref{WECh1}), the real energy density $\epsilon$ can be
seen to be positive at the horizon for a wide range of slices. It
can be shown that the parameter, ${M} \propto \mbox{ \{the
mass of the 5D bulk black hole\}}$, needs to be above a threshold
$M_0 \approx 0.15$ for all the negative values of
$\gamma$ to give a positive energy density at the horizon.
Unless otherwise stated, we will only consider such bulks in what
follows. For the dominant energy condition, we find that at the horizon,
$|\epsilon/p_\rho|=1$, and in fig.~(\ref{DECh3}),
we plot $|\epsilon/p_\theta|$ for a few values of $M$
from $1$ to $20$. For these, the dominant energy condition is satisfied
for $M$ small enough at least for ${\gamma} \in [-0.15,0]$,
as is illustrated by the plots. It can be shown that for large $-\gamma/l$,
$|\epsilon/p_\theta|<1$ (so the dominant energy condition is violated).
%ccforaki

\begin{figure}[t]\centering\includegraphics[angle=0,width=4in,height=4in]{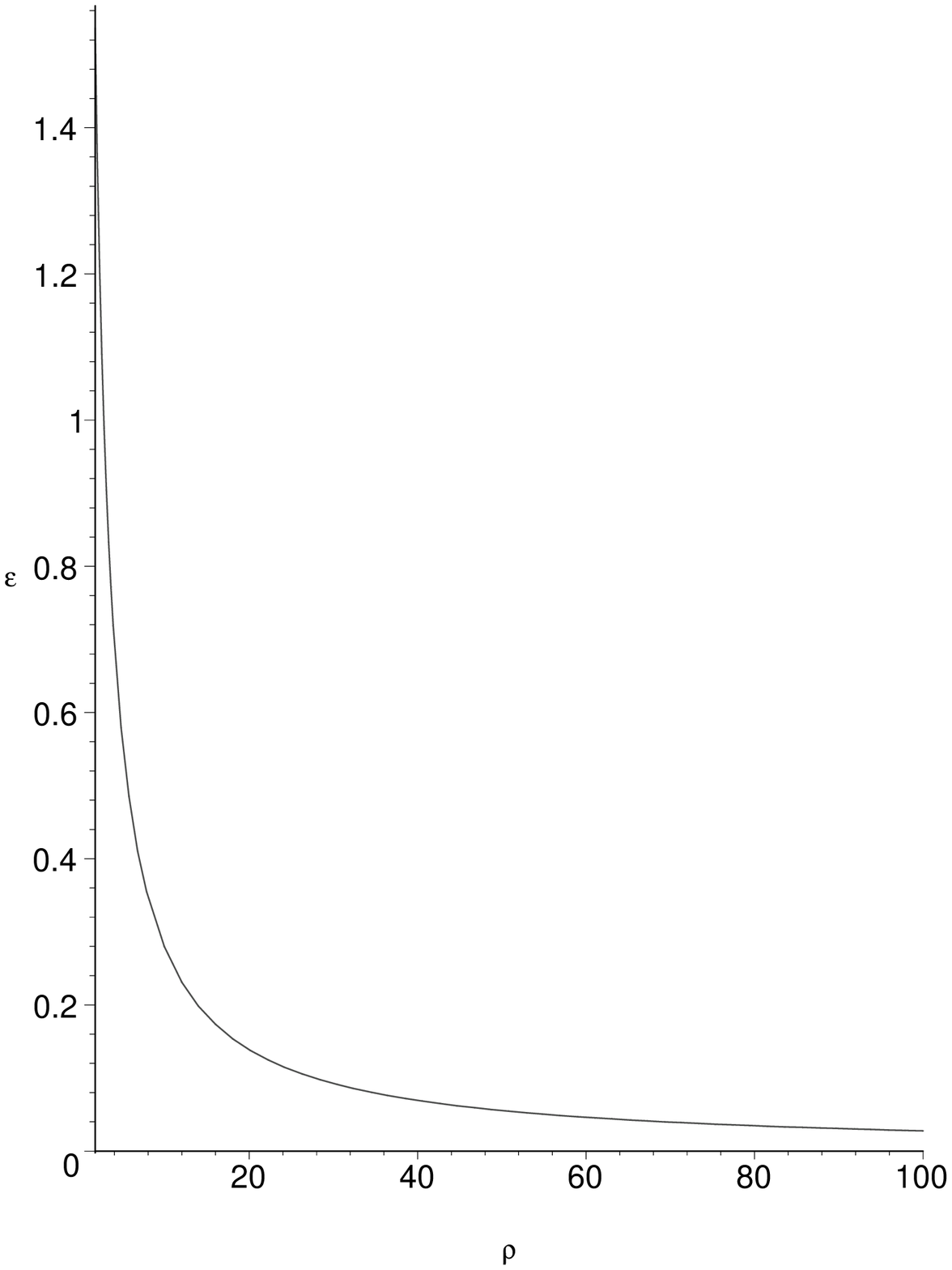}
\caption{The real energy density on the brane is seen to be positive everywhere outside the horizon. We have taken the particular values $M=2$
and $\gamma=-0.1$ but the qualitative behavior is the same for $0.2\lesssim M \lesssim 10$ and $0 \leq -\gamma \lesssim 0.15$. \label{WECe1}}\end{figure}

\begin{figure}[t]\centering\includegraphics[angle=0,width=4in,height=4in]{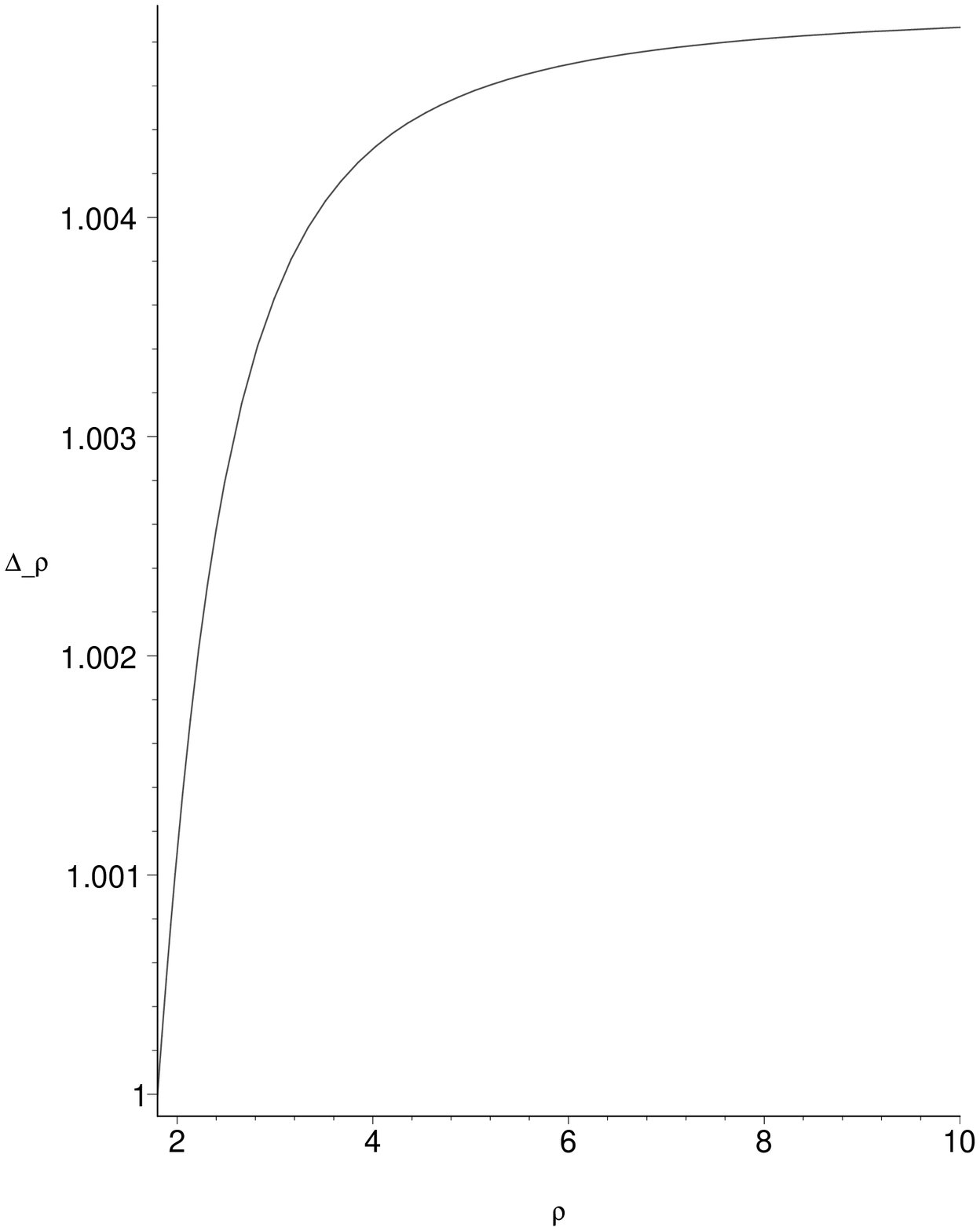}\caption{$\Delta_\rho=|\epsilon/p_\rho|$
on the brane is seen to be larger than 1 everywhere outside the horizon. We have taken the particular values $M=2$ and $\gamma=-0.1$
but the qualitative behavior is the same for $0.2\lesssim M \lesssim 10$ and $0 \leq -\gamma \lesssim 0.15$.
\label{DECe2}}\end{figure}

\begin{figure}[t]
\centering\includegraphics[angle=0,width=4in,height=4in]{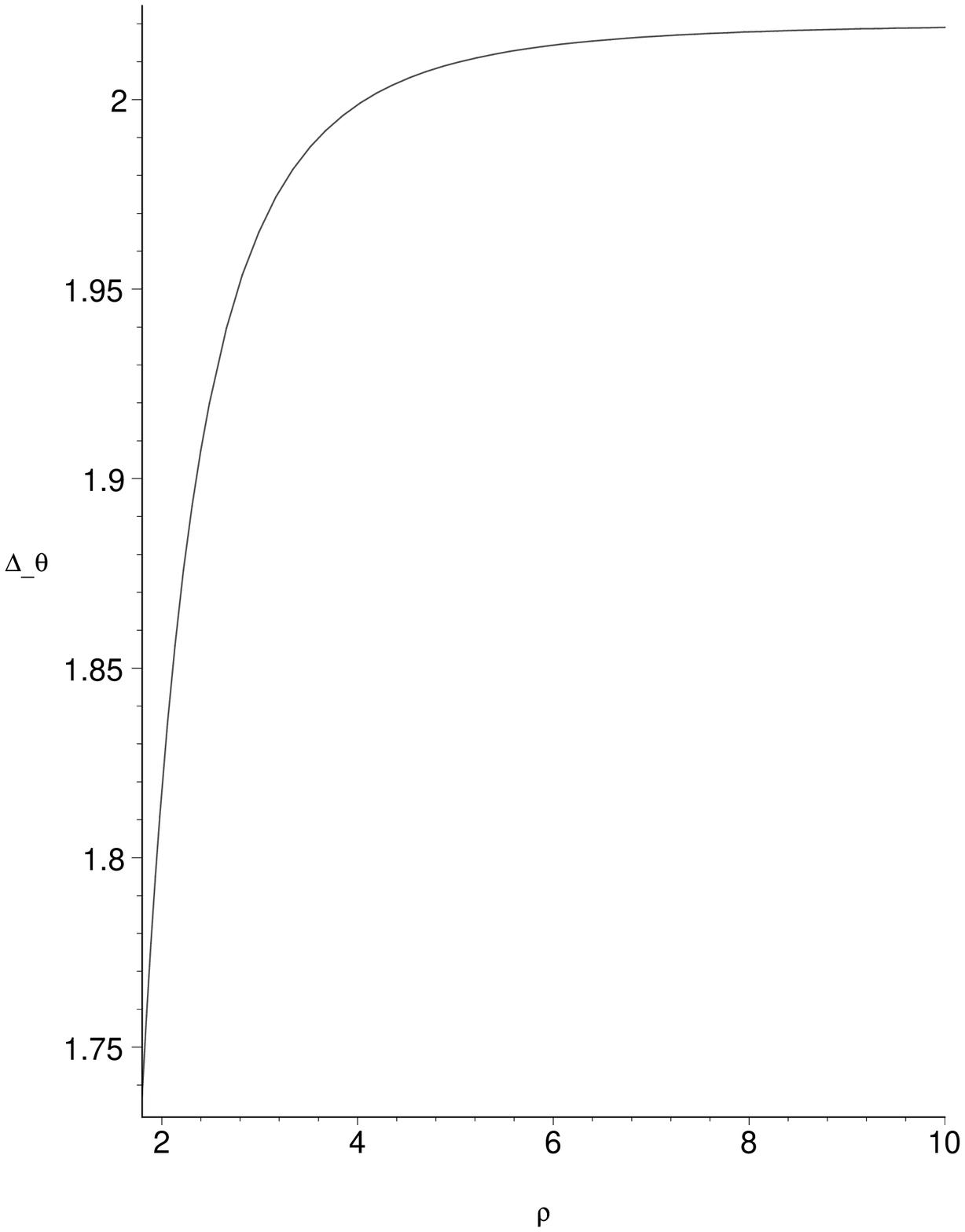}
 \caption{$\Delta_\theta=|\epsilon/p_\theta|$ on the brane is seen to be larger than
          $1$ everywhere outside the horizon. We have taken the particular
         values $M=2$ and ${\gamma} = -0.1$, but the qualitative behavior is the same for $0.2\lesssim M \lesssim 10$ and $0 \leq -\gamma \lesssim 0.15$.
\label{DECe3}}\end{figure}

For clarity, we now consider an explicit example of slice:
$M=2$ and ${\gamma} =-0.1$. The dominant energy condition
is satisfied everywhere outside the horizon $\rho_0 \approx 1.70$,
as is illustrated in figs.~(\ref{WECe1}, \ref{DECe2}, \ref{DECe3}).

Other values of $M$ and $\gamma$ could have been chosen.
For our particular choice, there is a singularity of type $a/$
at ${\rho} = {\rho}_a \approx 0.68$, where our brane changes signature.
Assuming the slice can be continued past this first singularity,
there is another curvature singularity (of type $c/$) at
${\rho} = {\rho}_c=0.1$.

To conclude this section we would like to see what happens asymptotically
for the effective stress tensor. We have
\bena
\label{effective WEC asympt}
\tilde{\epsilon}&=&\frac{4\gamma(\gamma^2+4l^2)}{l(2l^2+\gamma^2)^2}\frac{l}{\rho} +O \left(\frac{l^2}{\rho^2} \right) \,, \\
\tilde{p_\rho}&=&-\frac{2\gamma(\gamma^2+8l^2)}{l(2l^2+\gamma^2)^2}\frac{l}{\rho} +O\left( \frac{l^2}{\rho^2} \right) \, , \\
\tilde{p_\theta}&=&-\frac{8l^2\gamma}{l(2l^2+\gamma^2)^2}\frac{l}{\rho}
  + O \left( \frac{l^2}{\rho^2} \right) \,.
\eena
Until now, in order to have a positive vacuum energy on our brane,
we imposed on $\gamma$ to be negative. We here see that this constrains
the effective matter to violate the weak energy condition as soon as
$\rho$ is large enough. Had we chosen our brane to have a negative or
null vacuum energy, the effective matter would have had a positive
energy density asymptotically and it is easy to see that it would also
have satisfied the dominant energy condition. It is to be noted that
particular slices with $\gamma> 0$ can be found such that the real matter
satisfies the dominant energy condition everywhere outside
the horizon as before. Such an example is given
by $(M=2,\gamma=0.1)$.
The slice $\gamma=0$ for $k=-1$ or $0$ (generalizing the $k=1$
``equatorial slice") allows us to treat the problem analytically,
this will be the topic of the next section.
%-----------------------------------------------------------------------

\subsection{Black hole in asymptotically totally geodesic brane}

In this section, we concentrate on the $\gamma=0$ case, with $k\leq 0$,
which implies $\sigma=0$.
Since real and effective matter die off at spatial infinity
in the braneworld, the extrinsic curvature
of such a brane vanishes at spatial infinity. This makes the brane
asymptotically totally geodesic.

When $\sigma=0$, the real matter fields (i.e., source term linear
to $T_{\mu \nu}$) are decoupled from the brane gravity
(see eqs.~(\ref{G}) and (\ref{Teff})). In this case the only
source for the brane gravity through
eq.~(\ref{G}) are ${\cal S}_{\mu \nu}$ and ${\cal E}_{\mu \nu}$
tensors coming, respectively, from the junction conditions
and the projection of the 5D Weyl tensor onto the brane.
We incidentally note that this limit corresponds
in the holographic picture of \cite{Tanaka2,Casadio2,Germani,Emparan} to
the situation in which our black hole is surrounded by
``quantum'' matter only.

These slices provide explicit analytical examples of spherically
symmetric, localized brane black holes where the surrounding matter
does not violate dominant energy conditions anywhere outside the horizon.
In these black hole solutions, there is only one singularity which
is a central singularity hidden by a Killing horizon.
In fact, the induced metric on the brane is
\be
{}^{(4)}ds^2=- \left(k-\frac{2Ml}{\rho^2}+\frac{\rho^2}{l^2}\right)dt^2
     + \frac{d\rho^2}{k-\frac{2Ml}{\rho^2}
     + \frac{\rho^2}{l^2}}+\rho^2 d\Omega_{(2)}^2 \,, \quad
   \mbox{with} \,\,\,\, k= -1 \,, \,\,\, 0 \,.
\label{metric:asympto-geodesic}
\ee
So, the Killing horizon is located at
\be
  \rho_0=\frac{l}{2} \sqrt{-2k+2\sqrt{1 + 8M/l} } \,.
\ee
The matter content on the brane is
\begin{eqnarray}
\epsilon&=&\frac{2\sqrt{1-k}}{\rho} = - p_{\rho}
         = -2 p_{\theta} = -2 p_{\phi}\,\cr
\tilde\epsilon&=&(1-k)\frac{\rho^2-Ml}{\rho^4}=-\tilde p_{\rho}\ ,\cr
\tilde p_\theta&=&\tilde p_\phi=-(1-k)\frac{Ml}{\rho^4}\ ,
\end{eqnarray}
where the ``$M$'' contribution comes from the projection of the Weyl
tensor onto the brane and the Einstein tensor is diagonal.

In the case of big black hole mass $M\gg l$, it is noticeable that
the energy density for the effective matter, although positive
asymptotically, becomes negative when approaching the black hole horizon
(when $\rho \leq \sqrt{Ml}$).
%%%
%%%
This is a local effect,
and when we consider the effective matter to correspond to quantum matter
\cite{Tanaka2,Emparan,Casadio2,Germani}, this is reminiscent of
semi-classical considerations in asymptotically flat black hole spacetimes
\cite{Page,Visser}. The holographic interpretation can be
trusted only in this limit \cite{Tanaka2}.
%%%
It is easy to calculate the curvature scalars.
They are everywhere regular except at $\rho=0$. For instance,
\be
  R^\alpha{}_\alpha=-\frac{12}{l^2}+\frac{2(1-k)}{\rho^2}\ ,
\ee
and $R_{\alpha\beta}R^{\alpha\beta}\propto \rho^{-8}$,
$R_{\alpha\beta\gamma\delta}R^{\alpha\beta\gamma \delta}\propto
\rho^{-8}$ near $\rho=0$. We therefore have a central space-like
singularity located at $\rho=0$, provided $M>0$.
%%%
(When $k=-1$, even for $M<0$ the bulk spacetime possess
an event horizon that hides a time-like bulk singularity,
provided $-1 <8M/l$. Accordingly the brane's singularity also
is time-like, which is hidden inside the brane black hole horizon.)
%%%

\par %%%
The zero mass bulk black hole case, $M=0$, is special:
In this case, our bulk is locally pure $AdS_5$, hence is regular
everywhere.
With the coordinate choice $k=-1$, $AdS_5$ bulk has a (non degenerate)
Killing horizon at $r=0$. Our brane (\ref{metric:asympto-geodesic})
(with $M=0$, $k=-1$) then describes a black hole with a Killing horizon
at $r=\rho =l$. The central curvature singularity of this brane,
due to this non-trivial embedding, is space-like and hidden by
the Killing horizon.
On the other hand, the choice $k=0$ corresponds to the
horospherical coordinates of $AdS_5$ bulk, which now has a Killing
horizon at $r=0$ of this coordinate. Our brane in this case hits
this bulk Killing horizon at $\rho =0$, but now the brane becomes
singular on that horizon.
The dominant energy conditions are nowhere violated for both
effective and real matter.

%-----------------------------------------------------------------------

%-----------------------------------------------------------------------
\section{Conclusion}
\label{conclusion}
%-----------------------------------------------------------------------

One of the most important unsolved problems in braneworld scenario
is probably the missing localized black hole solution on a braneworld.
Until now, only negative results of (vacuum) localized braneworld black
holes had been put forward. Motivated by the conjecture that
{\it localized black hole on a Randall-Sundrum brane}
 holographically corresponds to a {\it semiclassical
four dimensional black hole} \cite{Emparan, Tanaka2},
we tried to find a non-vacuum, asymptotically AdS black hole solution
on such a (non totally geodesic) brane.

To achieve this goal, we hunted for possible slices of
the Schwarzschild-AdS bulk which cut the bulk black hole horizon
producing a smooth horizon on the brane. We showed that this is
possible if one introduces suitable matter on the brane,
solely determined by the junction conditions on the brane itself.
Requiring such matter to be realistic, we looked for slicing
corresponding to a brane filled by matter satisfying the dominant energy
conditions.

More explicitly, we studied the simplest one-parameter family of
slices which obeyed these constrains. Although our parameter turned out to
be constrained for our energy conditions to be satisfied
(outside the horizon), we found a whole range of values for which our slices
correspond to a (regular) localized ``black hole" on a brane.
In some particular cases, corresponding to a generalization of
the ``equatorial slice" of the spherical Schwarzschild AdS bulk
to hyperbolic and flat three-dimensional geometries, we found
explicit analytical solutions with a horizon hiding a single point-like
singularity at the center. We also noticed that for non-zero large bulk
black hole mass, the energy conditions for the effective matter are
satisfied outside a spherical region surrounding the black hole horizon.
This is reminiscent of semi-classical results obtained in asymptotically
flat black hole spacetimes (see \cite{Page,Visser} and references therein).
For zero mass bulk black hole case, both the real and effective
energy conditions are satisfied everywhere outside the horizon.

We also showed that it is possible for the part of
our braneworlds that is hidden by the Killing horizon
to undergo a signature change.
%%%
From the braneworld view point, the appearance of the Euclidean
signature region might be interpreted in the quantum theoretical
context, such as Euclidean quantum gravity on the braneworld.
On the other hand, the bulk spacetime is everywhere Lorentzian, hence
one may expect that the braneworld signature change could entirely be
understood in terms of bulk classical theory. Such an expectation is
in accord with the spirit of a holographic idea in the sense that
quantum phenomena on braneworld have some correspondence
to bulk classical phenomena.
%%%
However, whether such a signature changing braneworld can be
realized as a solution of a well-posed initial value problem
(e.g., 5D Einstein equations with suitable initial data)
is a non-trivial question.

By construction, our results fit in a Randall-Sundrum
braneworld scenario. We nevertheless believe that our solutions can
be of phenomenological importance beyond that framework
in understanding real astrophysical or microscopic black holes
if extra dimensions are part of an ultimate theory of gravity.

Finally, from the perspective of the conjecture of \cite{Emparan,Tanaka2}, we would like to conclude by pointing out that the brane black hole
solutions found in this paper are, to our knowledge, neither known
semi-classical solutions nor possibly obtained by perturbations
thereof (see \cite{Emparan2} for further similar comments).
We found that black holes satisfying energy conditions are possible only in the case $\Lambda_4\simeq \Lambda_5$.
In this regime, four dimensional gravity is not localized at least at spatial infinity \cite{karch}. Therefore, although
our results apparently contradict the conjecture of \cite{Emparan} and \cite{Tanaka2}, it is not clear whether our black hole solutions
can be used for this holographic conjecture \footnote{We thank Takahiro Tanaka for pointing this out to us.} as, at least at spatial infinity,
we would expect our black hole solutions to correspond to a deformed conformal field theory without gravity \cite{deformed}.
Clarification of this problem is beyond the scope of the present work.
%ccforaki

\bigskip

\begin{center}
{\bf Acknowledgments}
\end{center}
The authors are indebted with M.~Sasaki for enlightening discussions. We also wish to thank S.~Hartnoll, S.~Hawking, R.~Maartens and G.~Procopio
for useful comments and discussions. AI was supported by NSF
grant PHY 00-90138 to the University of Chicago. CG was supported in part
by JSPS fellowship (JSPS/FF1/412) and in part by PPARC research grant (PPA/P/S/2002/00208).

\end{document}